\def\ROSAT{{\it ROSAT\/\ }}
\def\ASCA{{\it ASCA\/\ }}

\def\Ir{IRAS~13224--3809\ }
\def\Irc{IRAS~13224--3809}

\def\ltsima{$\; \buildrel < \over \sim \;$}
\def\simlt{\lower.5ex\hbox{\ltsima}}
\def\gtsima{$\; \buildrel > \over \sim \;$}
\def\simgt{\lower.5ex\hbox{\gtsima}}

\documentstyle[psfig]{mn}

\title[\ROSAT monitoring of \Ir]
{\ROSAT monitoring of persistent giant and rapid variability in the 
narrow-line Seyfert 1 galaxy \Ir }

\author[Th. Boller, W.N. Brandt, A.C. Fabian \& H.H. Fink]
{\parbox[]{6.5in}{Th. Boller,$^1$ W.N. Brandt,$^{2,3}$ A.C. Fabian$^3$ and H.H. Fink$^1$}\\
\\
$^1$ Max-Planck-Institut f\"ur Extraterrestrische Physik, 85748 Garching, Germany\\
$^2$ Harvard-Smithsonian Center for Astrophysics, 60 Garden Street, Cambridge, Massachusetts 02138, USA\\
$^3$ Institute of Astronomy, Madingley Road, Cambridge CB3 0HA\\
}
 
\begin{document} 

\maketitle

\begin{abstract}
We report evidence for persistent giant and rapid X-ray variability
in the radio-quiet, ultrasoft, strong Fe~{\sc ii},
narrow-line Seyfert~1 galaxy \Irc.
Within a 30 day \ROSAT High Resolution Imager (HRI) monitoring 
observation at least five giant amplitude count rate 
variations are visible, with the maximum 
observed amplitude of variability being about a factor of 60.
We detect a rise by a factor of about 57 in just
two days. \Ir appears to be the most X-ray variable Seyfert known,
and its variability is probably nonlinear. We carefully 
check the identification of the highly variable
X-ray source with the distant galaxy, and it appears to be
secure. We examine possible explanations for the giant variability.
Unusually strong relativistic effects and partial 
covering by occulting structures on an accretion disc
can provide plausible explanations of the X-ray data, and 
we explore these two scenarios. Relativistic 
boosting effects may be relevant to understanding the 
strong X-ray variability of some steep spectrum Seyferts 
more generally.  
\end{abstract}


\section{Introduction}

\begin{figure*}
       \psfig{figure=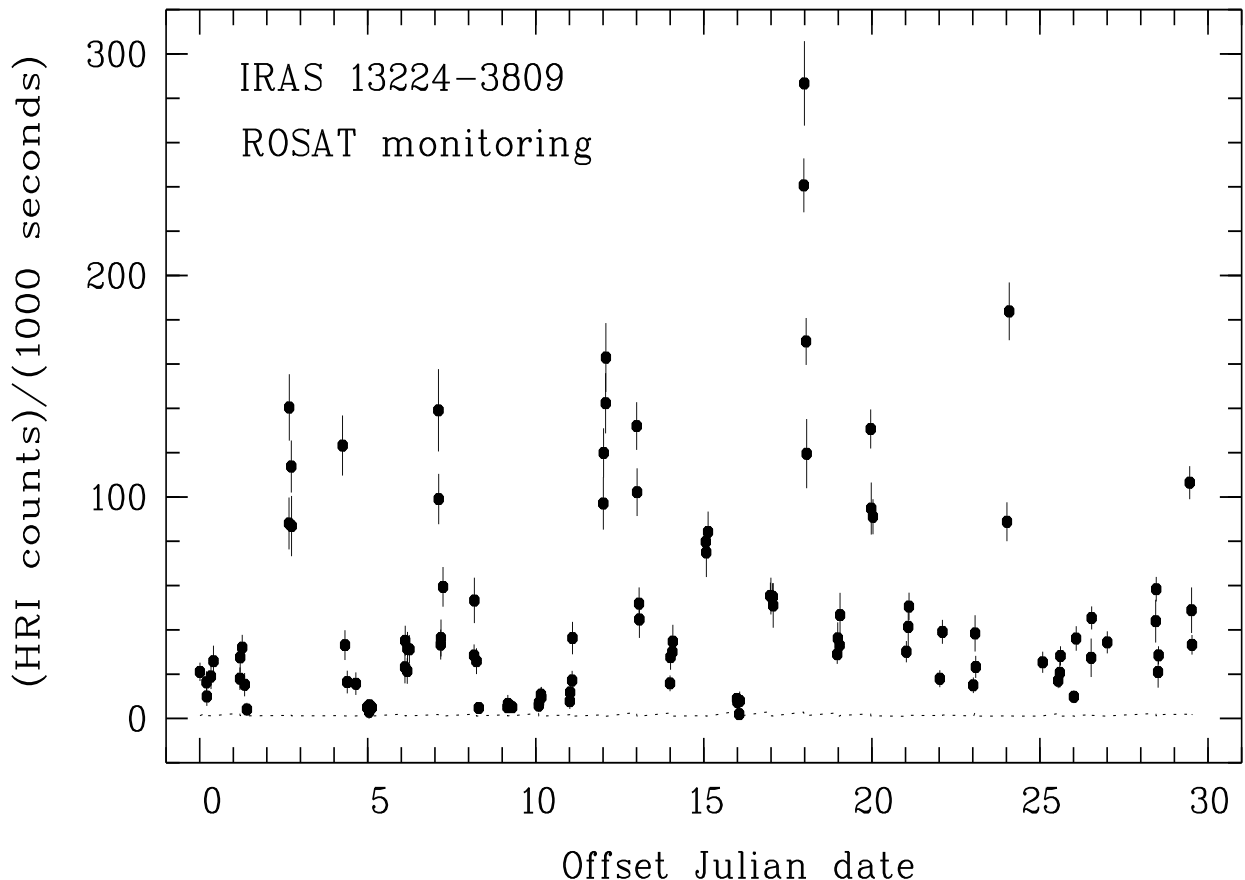,height=12.0cm,width=17cm,clip=}
      \caption{
\ROSAT HRI light curve for \Ir obtained during a 30 day monitoring observation 
between 1996 January 11 and 1996 February 9.
The abscissa label gives the Julian date minus 2450093.523 days.
Each data point is plotted at the middle of the exposure interval from 
which it was obtained, and the sizes of the exposure intervals lie
within the data points themselves.
The total exposure time is 111.313 ks, and the source is centered 
on-axis in the field of view.
The dashed curve indicates the background counting rate within the source
extraction circle as a function of time. 
At least five giant amplitude variations are visible (see the text for
details). We obtain similar results when ignoring HRI channels 1--3 and can
exclude any ultraviolet leak from having serious effects.
}
\end{figure*}

\noindent Variations of the continuum emission in Seyfert galaxies 
can give upper limits on the sizes of their emitting regions 
and the masses of their central black holes (e.g. Terrell 1967;
Kinman 1968; section IV of Zwicky 1971; 
Elvis 1976; Tananbaum et~al. 1978),
assuming that the variability is not affected by beaming 
or relativistic motions. Variations can also give 
clues about the physical processes operating in the black 
hole region (e.g. Lightman, Giacconi \& Tananbaum 1978; Fabian 1979;
in this paper the `black hole region' shall be taken to mean the
region within 50 Schwarzschild radii from the supermassive
black hole). Giant amplitude X-ray 
variability by more than a factor of $\sim 15$ 
appears to be fairly rare among Seyferts, although several 
dramatic examples of giant variability 
have recently been found in 
ultrasoft narrow-line Seyfert 1 galaxies (NLS1; also 
sometimes referred to as `I~Zw~1 class objects'). The 
NLS1 Zwicky 159.034 
(Brandt, Pounds \& Fink 1995; Grupe et~al. 1995a; Brandt et~al. 1996a), 
WPVS007 (Grupe et al. 1995b) and possibly
PHL 1092 (Brandt 1995; Forster \& Halpern 1996)
have been seen to have giant amplitude variability with time 
scales of the order of years.
The most peculiar object among the giant variable NLS1 class 
is perhaps \Irc, as amplitude variations of a factor of $\sim 30$ 
were once seen to occur during a single 80 ks \ASCA observation 
starting on 1994 July 30 (Otani, Kii \& Miya 1996).
\Ir was known before as the most violently variable Seyfert observed
with \ROSAT (Boller et~al. 1993; 
Boller, Brandt \& Fink 1996, hereafter BBF96).
The shortest doubling time in a \ROSAT AO-3 observation was 
$\approx 800$~s, and the maximum amplitude of variations 
was a factor of about 8 within approximately 2 days. 
The apparent efficiency derived from the variability was constrained 
to be greater than about 8 per cent using the arguments of Fabian (1979). 
This large efficiency exceeds the maximum Schwarzschild black hole accretion 
efficiency of 5.7 per cent, perhaps suggesting 
relativistic effects or accretion onto a Kerr black hole. 

In this paper we report the results of a \ROSAT HRI monitoring
observation of \Ir over a 
period of 30 days, in which we have found
the strongest persistent X-ray variability known in a 
Seyfert galaxy. In Section~2 we present the monitoring data.
In Section~3 we briefly establish the radio-quiet nature of
\Irc. Evidence for the association of the variable X-ray 
source with the distant galaxy \Ir is given in 
Section~4, and the discussion may be found in 
Section~5. Section~6 contains the summary.

We adopt $H_0=50$ km s$^{-1}$ Mpc$^{-1}$ and  
$q_0 = \frac{1}{2}$ throughout. The redshift of 
\Ir is $z=0.0667$ (Boller et~al. 1993), and the Galactic 
column at its position is $5.3\times 10^{20}$ cm$^{-2}$
(Stark et~al. 1992).  

\section{ROSAT HRI monitoring results}

\begin{figure}
       \psfig{figure=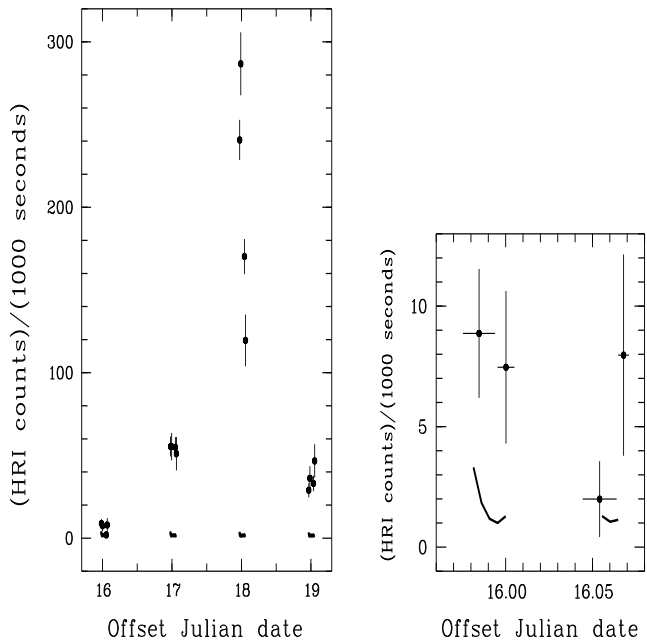,height=8.7cm,width=8.7cm,clip=}
      \caption{
\ROSAT HRI light curve for days 16 to 19. We also 
illustrate the low count rate levels we measure.  
The Julian date offset is as per Figure 1. 
The lower solid curves indicate the background counting rate within the 
source cell as a function of time. The count rate from \Ir rises 
by about a factor of 57 within two days.
}
\end{figure}

\begin{figure}
       \psfig{figure=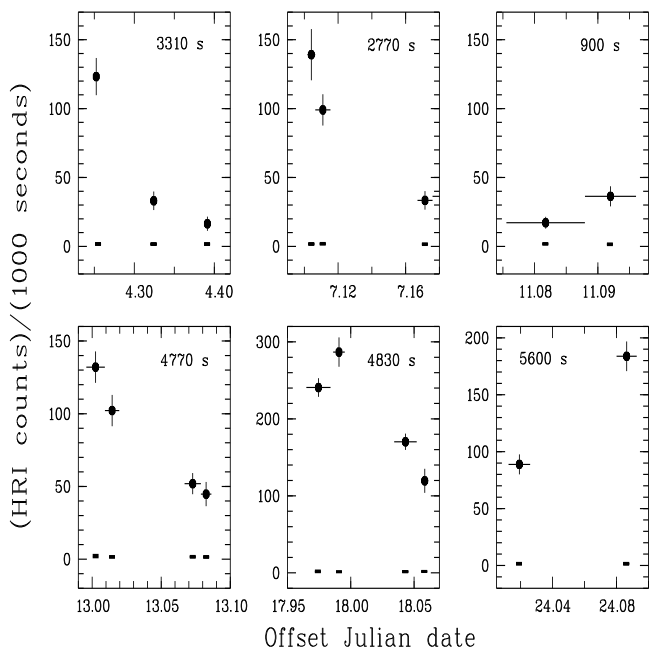,height=8.7cm,width=8.7cm,clip=}
      \caption{Shortest observed doubling times during our monitoring
observation. The integration time per bin is always larger than 400 seconds 
to avoid count rate modulation due to the \ROSAT wobble. Doubling times are 
given in the upper corners of the panels. The Julian date offset is 
as per Figure 1, and the lower solid curves indicate the background 
counting rate within the source cell as a function of time.}
\end{figure}

\subsection{Observations and the master image}

Thirty daily pointed observations using the HRI detector 
(David et~al. 1996) onboard the \ROSAT satellite (Tr\"umper 1983) 
were obtained for \Ir between 
1996 January 11 (0:33:42) and 1996 February 9 (13:23:28). 
The HRI is sensitive to photons 
between $\approx$ 0.1--2.4 keV.
The total exposure time was 111.313 ks, and the
source was centered on-axis in the field of view.
Daily exposure times range from 1.4--5.3 ks, allowing
several high quality count rate determinations per
day. Daily exposure times were often spread over a 
significant fraction of each observation day, allowing
a range of timescales to be probed. 

We made a master HRI image by adding the images from all
30 observations together. The centroid position 
for \Ir in the master HRI image,
computed from a Gaussian fit to the spatial distribution, is 
$\alpha_{2000} = \rm 13^h25^m19.59^s \pm 0.03s$,
$\delta_{2000} = \rm -38^o24^{'}55.45^{''} \pm 0.55^{''}$.
The internal HRI position error is of order 5 arcsec, and it
will be considered in more detail below when discussing the positional 
coincidence of the X-ray and optical counterparts.

\Ir appears to be slightly elongated in the master HRI image.
Gaussian fits to the radial distributions in the x-direction and the 
y-direction result in a ratio of the 
major axis radius ($\approx 8$ arcsec) to the minor 
axis radius of about 1.19 at half maximum intensity. 
HRI images of point sources often appear to be elongated on such scales,
and this has been attributed to errors in the attitude correction as 
investigated in detail by Morse (1994). 
This slight elongation does not affect the results presented below,
and \Ir does not show any evidence for extent that obviously cannot  
be attributed to slight attitude errors. 

\subsection{Source and background cells}

The source cell size was obtained from a radial profile of counts within
a ring centered on the centroid position of \Irc. 
At a distance of 20 arcsec from the centroid position 
the radial profile is, to within the errors, consistent with the background 
level. The number of source plus background photons within 20 arcsec from
the centroid position is $5797\pm 76$, corresponding to a mean 
density of 4.62 source and background photons per
arcsec$^2$ in the source cell (of course, the count density is not
uniform across the source cell).

The background was determined from a ring 
around the source centroid position
with an inner radius of 40 arcsec and an outer radius of 120 arcsec.
Within this ring there are no discrete X-ray sources, and the
radial distribution is constant to within the errors.
The number of background photons normalized to the source cell size 
is 195$\pm$14, corresponding to 0.16 background photons per arcsec$^2$. 
In all the following analysis we, of course, take into account the 
(significant) time variations of the background count rate. These
cannot materially affect the results below. 

\subsection{Observed giant and rapid variability}

A careful examination of the arrival times of the source photons 
shows very strong deviations from the mean count rate. 
In Figure 1 we illustrate the observed variability by showing a 
plot of the \ROSAT HRI count rate versus time. The data are binned 
into bins with widths of 2000 s or less. The amount of 
integration time per bin is always larger than 400 seconds, to 
obtain good statistics and avoid 
apparent count rate variations 
due to the \ROSAT wobble. We note that
count rate modulation due to the wobble is significantly less
important for the HRI than the 
Position Sensitive Proportional Counter (PSPC).
The background counting rate in the source cell is
very small ($\sim 2\times 10^{-3}$ count s$^{-1}$ but time
variable) due to the excellent spatial resolution
(which leads to a small source cell size). 
We have performed our analysis using two independent
software packages, and we obtain generally consistent results.

In order to determine a conversion factor 
between HRI count rate and luminosity, we use
the spectral shape that has been observed with the 
\ROSAT PSPC and {\it ASCA\/}. We 
deliberately do not use just a simple
power-law model to determine the conversion factor due 
to the very steep spectrum of \Irc. The simple
extrapolation of a steep power-law spectrum to low 
energies where absorption is important can 
lead to unphysically large, or at least highly uncertain,
inferred luminosities (compare section 2 of 
Forster \& Halpern 1996 and the more restrained 
analysis in chapter 4 of Brandt 1995
for an example of this effect). We instead 
use a model with a power law and a soft excess, 
and we note that the presence of a soft excess component is
suggested by the PSPC and \ASCA data. While we strongly
doubt that the soft excess of \Ir arises via thermal 
bremsstrahlung emission (see Boller et~al. 1993), 
a power law plus bremsstrahlung soft excess model
does give a reasonable fit to the PSPC data 
and a column density close to 
the Galactic one. We derive that  
1 HRI count s$^{-1}$ corresponds to 
an absorption corrected 0.1--2.4~keV flux of 
$\rm 1.15\times 10^{-10}\ erg\ cm^{-2}\ s^{-1}$.
Other models with soft excesses also give 
similar results (we, of course, require the absorption
columns in such models to be at least as large as the Galactic
value). For example, a power law plus 
two blackbody soft excess model
leads to 1 HRI count s$^{-1}$ corresponding to
$\rm 1.68\times 10^{-10}\ erg\ cm^{-2}\ s^{-1}$.
Using equation 7 of Schmidt \& Green (1983) 
and the cosmological parameters of Section 1, we find that 
1 HRI count s$^{-1}$ corresponds to an isotropic luminosity of 
$\rm 2.9\times 10^{45}\ erg\ s^{-1}$ in 
the \ROSAT band (for the bremsstrahlung soft excess
model). Of course, when we
use a constant conversion factor between HRI count rate
and isotropic luminosity we are assuming that there is no strong
continuum spectral variability in the \ROSAT band. 
Boller et~al. (1993) did not find strong spectral variability
in the PSPC data, although there was evidence for some weak spectral 
variability. Also, Otani (1995) and Otani et~al. (1996) found that
the temperature of their blackbody component
(which dominates the \ROSAT band flux) was almost constant 
despite large flux variations. We shall discuss this issue
further below.     

We have computed the lowest and highest count rates observed to determine  
the maximum amplitude of variability. The lowest count rate is
observed during day 5. We have fitted a 
constant model to these data and obtain a source count rate of 
$(4.7\pm 2.5)\times 10^{-3}$ count s$^{-1}$. 
The maximum observed count rate was $0.287\pm 0.019$ count s$^{-1}$,
and it was detected in a 791 second exposure interval
starting on day 17.9861 (see Figure 1).
The most probable maximum variability amplitude is a factor of 61.
We are not able to use the standard approximate variance formula 
for error propagation due to the fact that the error on the minimum 
count rate is not negligible compared to the minimum count rate itself, 
but maximum variability amplitudes in the range 37--139 are 
most likely (Cauchy distributed; see 
section 2.4.5 of Eadie et~al. 1971 for details).

The most extreme amplitude variation in a short time
occurs between days 16.0160 and 17.9861 
(see Figure 2).
We observe an increase of the count rate from 
$(5.0\pm 1.9)\times 10^{-3}$ count s$^{-1}$
(averaged over an exposure interval of 4656 seconds) to 
$0.287\pm 0.019$ count s$^{-1}$
(exposure interval of 791 seconds). This corresponds to an amplitude 
for giant variability of about 57 within two days.
The \ROSAT band isotropic luminosity rises from about 
$1.5\times 10^{43}$ erg s$^{-1}$ to about
$8.3\times 10^{44}$ erg s$^{-1}$. If the emission 
from \Ir is isotropic, then this variation 
of $\sim 8.2\times 10^{44}$ erg s$^{-1}$ is truly remarkable.  
It would be roughly equivalent to a typical Seyfert 1
like MCG--6--30--15 abruptly rising up in soft X-ray 
luminosity to become almost as powerful as a quasar.   

We detect rapid variability throughout our monitoring.  
The shortest doubling time observed is about 900 seconds, and 
doubling times with higher statistical significances are found 
on the order of a few thousand seconds (see Figure 3).
These are important for our discussion below because
they suggest that the emission we are monitoring originates
very close to the central supermassive black hole. 

We have attempted to use the limited spectral resolution of
the HRI to look for any strong spectral variability. We have
compared light curves made using HRI channels 2--6 and 7--15.
Although our constraints are weak, we find no evidence for
spectral variability. 

Our monitoring clearly shows that the giant variability seen by
\ASCA (Otani et~al. 1996) was not 
a unique event and that \Ir appears to show
{\it persistent} giant amplitude X-ray variability. We shall 
discuss physical explanations for this remarkable variability 
in Section 5. 

\subsection{Nonlinear variability}

The light curves of active galaxies appear to be either
stochastic or chaotic and are not well described by 
easily-predictable deterministic 
differential equations (see sections 3 and 4 of Vio et~al. 1992).
Stochastic variability is perhaps weakly favoured by the data
(see section 4.1.1 of Mushotzky, Done \& Pounds 1993).
In addition, it has recently come to light that the optical and 
X-ray light curves of some active galaxies are nonlinear. That 
is, they represent a nonlinear stochastic process
rather than a linear one (they could also perhaps be
chaotic in which case they must be nonlinear).
Nonlinear variability has been seen
in an optical light curve of the OVV quasar
3C345 (Vio et~al. 1991), and Green (1993) was the first 
(to our knowledge) to establish probable nonlinear X-ray
variability from an active galaxy (the NLS1 NGC 4051).
The `three minute flare' from the NLS1
PKS~0558--504 is another probable example
of nonlinear X-ray variability (Remillard et~al. 1991). 
Leighly \& Marshall (1996) have claimed nonlinear X-ray 
variability from 3C390.3, although details have not yet
been presented. 

Our light curves appear to have periods of
relative quiescence as well as `flaring' periods. Such
behaviour suggests that the X-ray variability of \Ir may 
be nonlinear, since linear processes are not able to 
produce sudden bursts of large amplitude (see 
sections 3 and 5 of Vio et~al. 1992).
We have used the method of Green (1993) to test for 
nonlinearity (see his chapter 2), and 
we find that the variability of \Ir is probably 
nonlinear. Green (1993) shows that a (positive
definite) time series is nonlinear if the ratio of
its standard deviation to its mean is larger than
unity. Using the data points in Figure~1, we find that 
the unweighted mean count rate is  
0.050 count s$^{-1}$, 
the weighted mean count rate is
0.020 count s$^{-1}$ and 
the standard deviation is 
0.052 count s$^{-1}$
(here we are using equations 
2-7, 5-6 and 2-10 of Bevington 1969).
Therefore, the observed variability is probably nonlinear. 
An important caveat, as discussed in section 2.4.2 of 
Green (1993), is that we are assuming that our sample
means and standard deviation are accurate representations of 
the true means and standard deviation. The strong variability 
in earlier \ROSAT and \ASCA data suggests that this assumption 
is reasonable, but further monitoring is needed before
this assumption can be taken as having been proven valid. 

Another way examining light curves which show periods of 
relative quiescence as well as `flaring' periods
is to examine asymmetry about the mean. We have 
compared the distributions of absolute
deviations from the mean for the sets of data points  
larger than and smaller than the mean. 
If we use a Kolmogorov-Smirnov test to compare
these two distributions, we find that they are
inconsistent with greater than 99 per cent 
confidence (i.e. there are fewer data points above
the mean but they are further from the mean value). 

Vio et~al. (1991) and Green (1993) discuss some physical 
implications of nonlinear variability. Nonlinear X-ray
variability from \Ir would suggest that its light curve
cannot be modelled as a linear superposition of independent
events, such as flares in a corona or spots on a disk. 
Of course, flares and spots that interact nonlinearly or
are affected by a nonlinear flux amplification process are
allowed and are entirely plausible (see section 5 of
Vio et~al. 1992 for an example of a nonlinear spot model).

\subsection{Lomb-Scargle periodogram}

While one would probably not expect simple X-ray 
periodicity in a Seyfert light curve, it is 
nevertheless important to look for any
hints of periodicity, quasi-periodicity or other 
interesting harmonic content. This is especially true
when remarkable variability such as that we see from
\Ir is detected. Interesting harmonic content could 
arise from long-lived spots on an accretion disc, as
discussed by Sunyaev (1973) and 
Abramowicz et~al. (1992). 

In order to search for signals in 
our (unevenly sampled) data, we have computed 
the Lomb-Scargle periodogram of the light curve shown in 
Figure 1. We use the algorithm
given in section 13.8 of Press et~al. (1992), and our result
is shown in Figure 4. As
our sampling is not even, there is no precise Nyquist 
frequency that corresponds to our sampling. As commented
above, data point spacings range from about 1000 s to
about one day, with several count rate measurements
per day. In total, we have 108 data points spread over
30 days. If our sampling were even, the Nyquist frequency
would be about $2.1\times 10^{-5}$ Hz, and we would be
able to properly search for 
signals down to about 13.3 hours. However, as discussed in 
Scargle (1982), unevenly sampled data removes
some aliasing ambiguity. We have investigated signals
down to about 6.7 hours. We have kept in mind the 
issues discussed by Scargle (1982), and we recognize that 
the interpretation of power spectra made 
from nonlinear light curves requires 
caution (see Vio et~al. 1992). 

\begin{figure}     
      \psfig{figure=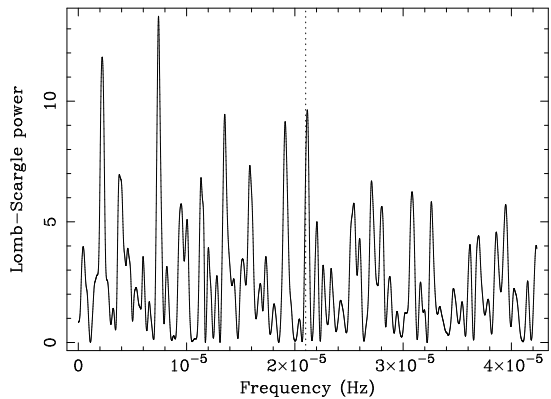,height=6.0cm,width=8.0cm,angle=270}     
      \caption{
Normalized Lomb-Scargle periodogram for the light curve
shown in Figure 1. The dotted vertical line shows the 
effective Nyquist frequency for even 
sampling (see the text). Note that peak significance
does not scale linearly with peak height, but is generally
a much stronger function of peak height (see Press et~al. 1992
but note also that there is an underlying red noise continuum). 
}
\end{figure}

The maximum value of the periodogram is found at 
$7.41\times 10^{-6}$ Hz, which corresponds to a period
of 1.56 days. This is below the effective
Nyquist frequency for even sampling given above.
It is difficult to rigorously assess 
the statistical significance of this maximum as 
it is superimposed on a poorly defined red noise power 
spectrum (the `false alarm' test in Press et~al. 1992 is 
clearly not appropriate). In addition, it must be 
remembered that our periodogram will be distorted at
some level by the window function of our sampling
pattern. It is beyond the scope of this paper to 
address these difficult issues in detail
(a more sophisticated analysis will be presented in
future work), but we show two cycles of a 
folded light curve in Figure 5. While we make no formal 
claim regarding periodicity, the folded light curve
may suggest some type of crudely periodic phenomenon.
The orbital timescale near a supermassive black hole is of 
order the timescale from the Lomb-Scargle peak.
This behaviour can be investigated further in the future using 
data which determine the underlying red noise continuum 
better. The count rate is seen to be persistently low during 
part of the putative cycle, while at other times it can be 
either high or low (i.e. if there is any modulation it is
of the count rate `envelope'). It appears unlikely that
we have missed any strong and simple periodicity of the
giant variability if it has a frequency of less
than $\sim 4\times 10^{-5}$ Hz. We have performed
Monte Carlo simulations using our sampling pattern and
measurement errors to see if we would have detected 
periodicity from a single, large, orbiting, X-ray hotspot
which causes the bulk of the giant variability (e.g. see
figure 5 of Bao 1992). It appears likely that periodicity
of this type would have been detected if it were present
(of course, some types of multiple hotspot models are entirely 
consistent with our data). The second highest Lomb-Scargle peak 
is at $2.21\times 10^{-6}$ Hz, which corresponds to a period
of 5.23 days. Folding the light curve on this period
gives a light curve qualitatively similar to the one
in Figure 5 in that, if anything, the count rate
envelope is modulated. 

\begin{figure}     
      \psfig{figure=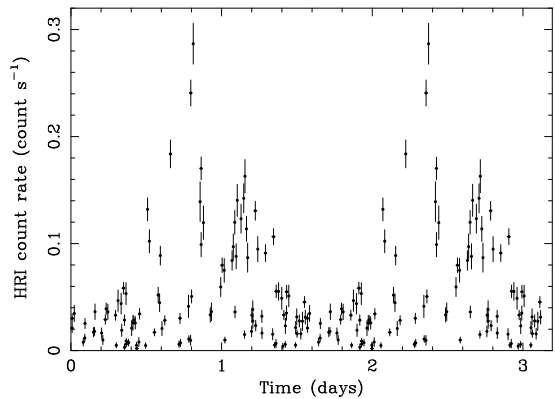,height=6.0cm,width=8.0cm,angle=270}     
      \caption{
Light curve for \Ir folded on the 1.56 day period of the text. Two
cycles are shown. 
}
\end{figure}

\subsection{Safety checks}

We have examined light curves for other sources in the HRI field of 
view (5 faint sources are detected, see Table 1).
The detection likelihood of the faint sources ranges
between 15 and 77 (see Cruddace et~al. 1988 for a discussion of
likelihood applied to \ROSAT source detection). A detection 
likelihood of 15 corresponds roughly
to 5 Gaussian sigma. While the faintness
of the other sources prevents detailed comparison, they do not
show evidence for correlated variability with \Irc.
We have checked the background light curve, and it is not
synchronized with the \Ir count rate.
At the edge of the HRI field of view there is the expected  
emission from a hot pixel (see David et~al. 1996). The hot 
pixel is very far away from
\Irc, and the hot pixel count rate is not correlated 
with the count rate of \Irc. As giant variability from 
\Ir has also been seen by {\it ASCA\/}, 
the variability we observe is almost certainly not due 
to a new hot pixel in the center of the HRI field of view 
or any other sporadic HRI effect.

\section{Radio-quiet nature}

In this section we briefly establish the radio-quiet 
nature of \Irc. This has not been done clearly in the 
literature, and it is relevant to our discussion section.

The 4.86 GHz radio flux density of \Ir measured at the VLA is 
$2.2\pm 0.3$ mJy = 
$(2.2\pm 0.3)\times 10^{-26}$ erg cm$^{-2}$ s$^{-1}$ Hz$^{-1}$  
(J. Condon and M. Dennefeld, private communication). In 
making this measurement C configuration was used, and the
beam size was 10 arcsec by 5 arcsec (with a position
angle of 0 deg). The optical $B_{\rm J}$ 
photographic IIIaJ+GG395 magnitude from the 
{\it Hubble Space Telescope} (HST) 
Guide Star Catalogue (GSC) 
is $13.43\pm 0.39$ 
(Lasker et al. 1990; Russell et al. 1990; Jenkner et al. 1990; see
the next section for further HST GSC information about \Irc).
Following section 3.1 of Hook et~al. (1994) we expect
a $B$ magnitude of about 13.5. 
Using equation 2 of Schmidt \& Green (1983) we derive an
optical flux density of about 
$1.8\times 10^{-25}$ erg cm$^{-2}$ s$^{-1}$ Hz$^{-1}$.  
The ratio of radio to optical flux density is about $R=0.1$. 
\Ir is therefore a radio quiet object using the formalism 
of Kellermann et~al. (1989, see their figure 4). The low
$R$ parameter for \Ir suggests that, even if it has
relativistic jets (see Falcke, Patnaik \& Sherwood 1996),
there is no reason to think that they are preferentially
pointed in the direction of Earth (as compared to
any other Seyfert~1 galaxy).

We furthermore note that \Ir was not detected in the radio
observations of Norris et~al. (1990), supporting its
radio-quiet nature (see their table 3). 

The optical polarization properties of \Ir are not known. In 
light of the remarkable X-ray variability and peculiar 
spectral energy distribution 
(see figure 3 of Mas-Hesse et~al. 1994), polarization 
measurements would be interesting. 

\section{Evidence for the association of the highly variable 
X-ray source with the distant galaxy \Ir }

Due to the fact that the variability we see from \Ir is
peculiarly strong, we have performed 
a careful examination of the 
identification of the X-ray source using all of the currently 
available data. While it is always difficult to formally prove, in
a strict mathematical sense, that there could not be a 
confusing source, below we show that there is no evidence
for such a source and that this possibility is very unlikely. 

\subsection{Position checking}

In this section we examine the $\approx 5$ arcsec \ROSAT HRI
error circle and its contents. 

The HST GSC gives a non-stellar classification at 
$\alpha_{2000}=\rm 13^h25^m19.28^s$, 
$\delta_{2000}=\rm -38^o24^{'}53.5^{''}$
(ID 0778700931; see the previous section for the $B_{\rm J}$ 
photographic magnitude). This position is 4.1 arcsec
from the HRI centroid position and lies within the X-ray 
error circle. Non-stellar 
classifications in the HST GSC can be given to
galaxies, blended objects and plate defects. 
If we carefully scrutinize an optical CCD image 
of \Ir taken with the La Silla 3.6 m telescope on 
1992 December 24 (see section 3 of Boller et~al. 1993), we 
see that the only visible object 
within the $\approx 5$ arcsec radius HRI error circle 
has the optical morphology of a galaxy. 
This is also true if we examine the actual UK Schmidt 
plate (which has significantly better resolution than the 
optical image available from {\sc skyview} at 
Goddard Space Flight Center). This morphological 
classification is confirmed by the optical spectra described 
below. We have carefully looked for other optical sources in 
the HRI error circle, and no other sources are visible in
the available images. The second nearest source found in
the HST GSC (ID 0778700947) is 59.2 arcsec from the HRI 
centroid position of \Irc.

The second nearest optical source (on the La Silla image
and the UK Schmidt plate) to the X-ray centroid
position is a star located at  
$\alpha_{2000}=\rm 13^h25^m18.7^s$, 
$\delta_{2000}=\rm -38^o24^{'}53^{''}$. It is about 10.7
arcsec from the X-ray centroid position and thus lies
significantly outside the X-ray error circle. While this
object is unlikely to be the counterpart
purely on positional grounds, we examine it further below. 
The third nearest optical source to the X-ray centroid
is about 18 arcsec away.  

It is sometimes possible to improve or check the HRI astrometry 
using serendipitous X-ray sources that have counterparts at 
other wavelengths with precisely known positions. We have
investigated other X-ray sources in the field of view for
this purpose (see Table 1). Five other X-ray 
sources are detected. While these
sources do not have NASA Extragalactic Database (NED) or 
Set of Identifications, Measurements and Bibliography for
Astronomical Data (SIMBAD) identifications, two of them appear to
have HST GSC counterparts. The right 
ascension offsets between the HRI and HST GSC positions are 
$\Delta\alpha = -6.2^{''}$ for RX~J~132240$-$3814 and
$\Delta\alpha = +0.0^{''}$ for RX~J~132459$-$3826.
The declination offsets are 
$\Delta\delta = +1.1^{''}$ for RX~J~132240$-$3814 and
$\Delta\delta = +2.8^{''}$ for RX~J~132459$-$3826.
The low number of coincidences and the offset
dispersion does not allow a precise determination of the 
internal HRI positional offset, although the offset 
vectors for \Ir and RX~J~132240$-$3814 are about the
same (suggesting that any offset correction would
probably further improve the already good agreement between 
the HRI centroid and the HST GSC position for \Irc). 

Furthermore, the X-ray position is 
constant to within the expected errors 
when considered as a function of source count 
rate (see Figure 6). That is, there is no strong 
evidence for a systematic position offset
when \Ir appears bright (as might occur if there 
were a nearby contaminating source). 

We note that the current positions given for \Ir in  
NED and SIMBAD are relatively poor and that 
our improved positions here supersede earlier ones 
obtained from detectors with poorer spatial 
capability. Furthermore, to avoid confusion, we
note that in the currently available {\sc skyview} 
images \Ir and the nearby star mentioned above 
appear to overlap due to the relatively poor 
digitization resolution.      

\begin{figure}
       \psfig{figure=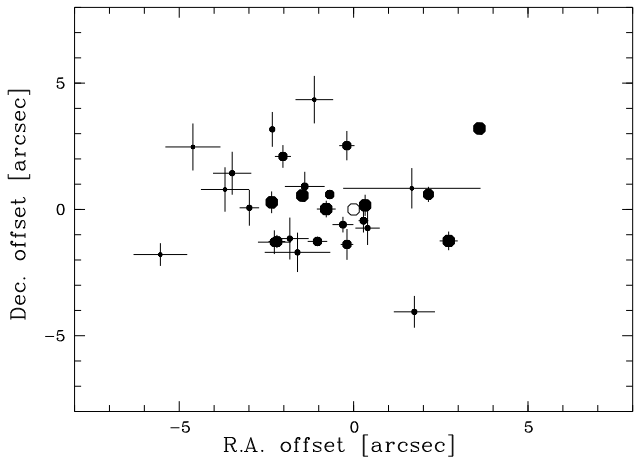,width=8.7cm}
      \caption{
Centroid positions for individual pointings at \Ir (filled hexagons) 
compared to the centroid position from the master image 
(open circle). The sizes of the filled hexagons are correlated to 
the average count rates of the individual pointings.
The centroid positions for \Ir in the individual pointings
were computed by fitting a Gaussian to the distribution in both the 
$\alpha$ and $\delta$ directions. There is a systematic error 
between the individual pointings of the order of 5 arcsec as suggested 
from the distribution of the centroid positions. There is no strong 
evidence for a systematic position offset when \Ir appears bright.
}
\end{figure}

\begin{table*}
{\scriptsize
  \begin{tabular}{|r|r|r|r|r|r|r|r|}
    \hline
(1)    & (2)               &  (3)                & (4)         & (5)      &  (6)              &    (7)           & (8)         \\
\ROSAT & HRI               &  HRI                & HRI count   & HST GSC  &  HST GSC          & HST GSC          &             \\
name   & $\alpha_{2000}$   &  $\delta_{2000}$    & rate        & ID       &  $\alpha_{2000}$  & $\delta_{2000}$  & p$_{\rm e}$ \\
\hline
\Ir                &  $\rm13^h25^m19.59\pm 0.03^s$ & $\rm-38^{o}24^{'}55.45\pm0.55^{''}$ &  ---                 & 0778700931   &    $\rm13^h25^m19.28^s$  &  $\rm-38^{o}24^{'}53.5^{''}$ & 0.4  \\
RX~J~132424$-$3811 &  $\rm13^h24^m23.94\pm 0.80^s$ & $\rm-38^{o}11^{'}19.19\pm8.25^{''}$ &  0.00131$\pm$0.00011 & ---          &    ---                   &  ---                          & ---  \\ 
RX~J~132440$-$3814 &  $\rm13^h24^m39.60\pm 0.40^s$ & $\rm-38^{o}14^{'}04.08\pm6.00^{''}$ &  0.00130$\pm$0.00011 & 0778700700   &    $\rm13^h24^m39.19^s$  & $\rm-38^{o}14^{'}03.0^{''}$ & 0.6  \\
RX~J~132459$-$3826 &  $\rm13^h24^m59.20\pm 0.40^s$ & $\rm-38^{o}26^{'}20.00\pm6.00^{''}$ &  0.00046$\pm$0.00006 & 0778701660   &    $\rm13^h24^m59.20^s$  &  $\rm-38^{o}26^{'}17.2^{''}$ & 0.4  \\
RX~J~132534$-$3828 &  $\rm13^h25^m33.70\pm 0.40^s$ & $\rm-38^{o}28^{'}50.00\pm6.00^{''}$ &  0.00036$\pm$0.00006 & ---          &    ---                   &  --- & ---  \\
RX~J~132537$-$3825 &  $\rm13^h25^m37.09\pm 0.40^s$ & $\rm-38^{o}25^{'}44.90\pm6.00^{''}$ &  0.00050$\pm$0.00007 & ---          &    ---                   &  --- & ---  \\   
    \hline
\end{tabular}
}
   \caption{Sources detected in the HRI field of view and possible
            optical counterparts from the HST GSC.
            As IRAS 13224$-$3809 shows giant amplitude variations (see the text),
            we do not give HRI count rate information here. The other 
            sources in the field of view maintain roughly constant HRI count rates,
            and the count rate errors are for 68.3 per cent confidence.
            Count rates are corrected for vignetting.  
            The last column ($\rm p_{\rm e}$) contains the positional 
            uncertainties for the HST GSC positions in
            arcsec. The astrometric errors of 
            HST GSC objects are discussed in Russell et al. (1990). 
           }
   \end{table*}

\subsection{Investigation of the nearby star}

In this section we examine the properties of the star that lies about 10.7
arcsec away from the X-ray 
centroid. While we first of all stress that this object
is very unlikely to be the counterpart purely on positional grounds (see
above), we have investigated it to be doubly cautious. Figure 7
shows an optical spectrum of the star taken on 
1992 December 24 with the La Silla 3.6 m telescope using 
the EFOSC spectrograph (A. Caulet, private communication). 
The Balmer absorption lines H$\alpha$, H$\beta$, H$\delta$ and H$\gamma$
are strong, but there are no He lines which one expects in early type 
stars. There is no strong Mg b feature around 5200 \AA, and
there is no evidence for emission lines as would be expected
from an accreting cataclysmic variable. 
The star appears to be a perfectly ordinary late G or
early K type star. 

\begin{figure}
       \psfig{figure=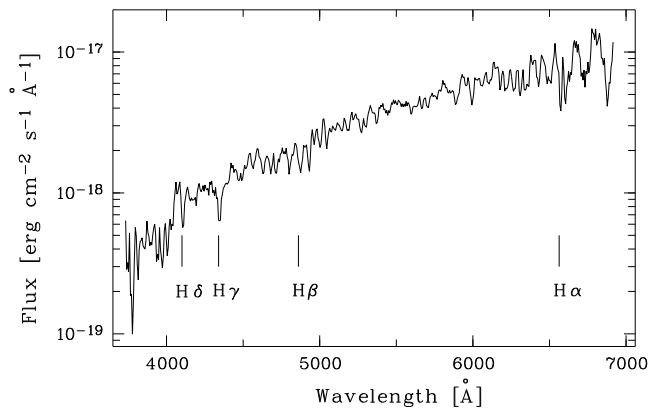,width=8.7cm,clip=}
      \caption{
Optical spectrum taken with the La Silla 3.6 m telescope of the 
star 10.7 arcsec from the X-ray centroid. 
}
\end{figure}

The flux ratio between \Ir and the nearby star, obtained from the
La Silla image, is about 6. This gives a visual magnitude for 
the nearby star of about 15.4. Using the nomograph shown in
figure 1 of Maccacaro et~al. (1988), we see that the combined 
X-ray flux and optical magnitude do not agree well with
those expected from an ordinary late G or early K type star. 
This further strengthens the idea that this star cannot be
the counterpart.    

\subsection{Optical spectra, ultraviolet spectra, 
X-ray spectral trends and X-ray variability trends}

The published optical 
spectrum of \Ir reveals that it is a NLS1 with strong
optical Fe {\sc ii} multiplets (see figure 3 of Boller et~al. 1993). 
C. Reynolds and M. Ward (private communication)
have independently taken a spectrum
at the Anglo-Australian Telescope on April 10, 1995 which confirms 
the redshift and the active nature of \Irc. Given the Seyfert 1
character of \Irc, its redshift and its emission in other
wavebands, it would be surprising if it were not an X-ray source.   

Further evidence against an unrelated contaminating source comes
from the remarkable Ly$\alpha$ line core variations reported from \Ir
by Mas-Hesse et~al. (1994). The variable Ly$\alpha$ line suffers
the expected cosmological redshift, leaving little doubt that
it is indeed from the distant galaxy. Interesting Ly$\alpha$ line
core variability has also been seen from 
the NLS1 Mrk 478 (see section 3.2 of Gondhalekar et~al. 1994). 

\Ir falls properly into place along the previously known 
distribution of H$\beta$ FWHM versus \ROSAT photon index 
(see figure 8 of BBF96), and this fact tends to support
the optical/X-ray match. An \ASCA observation of
\Ir shows a hard tail in its spectrum (see Otani et~al. 1996), 
as would be expected from an active galaxy but not from some 
other classes of X-ray emitting objects. The hard tail appears 
to vary in a synchronized way with the soft component. 

While persistent giant X-ray variability like that 
we report from \Ir is unprecedented,
the observed variability does appear to be an extension of
some previously known active galaxy variability trends. 
Other ultrasoft NLS1 with strong Fe {\sc ii} emission, such 
as WPVS007 and PHL 1092, have also shown very large amplitude
as well as rapid X-ray variability (see Section 1). 
NGC 4051, another NLS1 (see BBF96 for the NLS1 
character of this object),
shows very strong X-ray variability (e.g. figure 1 of
McHardy et~al. 1995 and Green et~al., in preparation). More 
generally, steep energy spectrum
active galaxies have been argued to possibly be the most
variable (see section 3.4.2 of Green, McHardy \& Lehto 1993). 
Evidence for relativistic X-ray beaming has been 
seen from PKS~$0558-504$, a peculiar ultrasoft NLS1 type 
object that is fairly radio loud (Remillard et~al. 1991). 
 
We do not see any strong periodicity in our light curve that is 
obviously suggestive of a contaminating source (see Figure 4).
This is true even if we look up to 4 times the effective
Nyquist frequency for even sampling. 

\section{Discussion}

\subsection{Physical interpretation of the observed variability }

The persistent giant amplitude X-ray variability we see from \Ir
suggests that interesting physical processes are operating in the
core of this galaxy. In this section we examine possible 
explanations for the variability and confront them with the
observations. We shall work within the supermassive black
hole paradigm for Seyferts. Starburst models
do not appear to be able to explain the rapid variations
shown in Figure~3 in conjunction with the giant variations
shown in Figure~1 and Figure~2. If the emission 
from \Ir is isotropic, we estimate that its black 
hole mass is larger than about $6\times 10^6$ M$_\odot$ using the
Eddington limit. If its Eddington fraction is 0.2, the black hole
mass is about $3\times 10^7$ M$_\odot$ and the black hole light
crossing time is $\sim 300$ seconds. We see rapid variability on
timescales comparable to this throughout our monitoring. 

We start by considering possible explanations which
involve processes far from the black hole, and we then 
successively move closer to the black hole. 
While we critically examine several possibilities 
for the variability below, 
the partial covering described in Section 5.1.3 and
the relativistic effects described in Section 5.1.7 
appear to be the most plausible possibilities, and
we examine these in the most detail.   


\subsubsection{Gravitational lensing by intervening objects}

Gravitational lensing by intervening objects (e.g. stars) is probably 
not a dominant factor in making Seyfert variability 
in general (see Abramowicz 1986), although it still could be 
relevant in some cases (e.g. Schneider \& Weiss 1987). 
There is no evidence for an intervening galaxy, so objects 
either in \Ir itself or the Milky Way would presumably be doing 
the lensing. 

Given the results of recent microlensing searches, 
strong lensing of an active galaxy as bright as \Ir by an 
object in our galaxy is very improbable. There are 
only a few thousand galaxies in the sky as bright as or brighter 
than \Ir (e.g. Sandage \& Tammann 1987). In addition, the 
observed microlensing events are seen to be much
slower than the variability we observe from \Ir (see 
equation 30 of Narayan \& Bartelmann 1997). Motion or
variability of the X-ray source itself 
could perhaps remove this problem and 
allow multiple giant variability events to occur, 
but in this case the giant variability could only persist for
as long as the length of the lensing event. Giant variability
has now been seen from \Ir both in 1994 July and 
1996 February, and this total span of time is longer than
observed large amplitude lensing events.  

Similar probability and timescale problems apply 
for objects in \Ir itself.


\subsubsection{Ionized absorption}

While \Ir may have an ionized (`warm') absorber, 
warm absorber changes alone cannot explain the observed 
variability. In order 
to cause the multiple rapid variability events shown 
in Figure 3, any putative warm absorber would have to either 
(1) rapidly move across the emission region several times or
(2) rapidly ionize and recombine several times. 
In the first case very high velocities would be needed
and the warm absorber would probably have to be located in 
the black hole region. \ASCA observations show that the typical 
warm absorbers seen from a number of bright nearby Seyfert galaxies 
do not appear to be located in the black hole region (e.g. their edges 
do not appear to be strongly blurred by Doppler effects), although 
there is no obvious a priori reason why ionized absorbing material
could not be located in this region. In the second case a
density of $\sim 1\times 10^9$ cm$^{-3}$ would be required
for the recombination timescale to be comparable to the
variability timescale. Material would not have to be in
the black hole region to have this density, but we note 
that the mechanism for causing the ionization change is
difficult to imagine. Typical warm absorbers are usually
thought to {\it respond to\/} rather than {\it control\/} 
changes in the observed flux from the central engine (i.e. they 
only affect the observed flux in a secondary rather than 
a primary way).

Extremely large edge depth changes would be 
required to cause giant amplitude variability
with a warm absorber. While \Ir has shown some spectral 
variability, it does not appear to be this extreme or of 
this type.   


\subsubsection{Partial covering}

No cold absorption changes have been observed from
\Irc, so time-variable partial covering by Compton thin matter
is unlikely. Time-variable partial covering by Compton 
thick matter (e.g. a non-axially symmetric elevation of 
the accretion disc surface) is possible 
because it could modulate the observed flux but not alter the 
spectral shape. In this scenario, the true HRI band 
luminosity of the source 
would persistently be $\simgt 8\times 10^{44}$ erg s$^{-1}$, and 
flux modulations would be due to a time-variable partial covering
fraction. For comparison, the 40--120 $\mu$m luminosity of \Ir is
$\sim 1.5\times 10^{45}$ erg s$^{-1}$. This and the rest
of the spectral energy distribution are not inconsistent
with a possible HRI luminosity of 
$\simgt 8\times 10^{44}$ erg s$^{-1}$ if \Ir is an
exceptional object (compare the spectral energy distribution
shown in figure 3 of Puchnarewicz et~al. 1995). 
We shall not discuss all the mechanical details of a partial
covering model here, as much of the discussion in section 5.1.4 of 
Done et~al. (1992) is directly applicable. 

Any partial coverer would have to be able to cross a significant
fraction of the emission region in a few thousand seconds if it 
causes the rapid variability (see Figure 3 and note that rapid
variability is seen even when the source is relatively bright),
and thus it would probably be in the black hole region.
It would have to avoid being sheared apart by the strong
angular velocity gradient in this region. We 
cannot use ionization parameter arguments like those in 
section 3.2.1 of Brandt et~al. (1996b) to argue against partial 
covering in the black hole region, as we do not have a strong 
upper limit on the mass of the accreting black 
hole (larger masses allow for lower Eddington fractions and lower 
ionization parameters). A thick partial coverer would probably be able
to remain at about the same ionization state as the part of
the disc local to it. Additionally, the observed
variability could be due to a combination of partial 
covering and intrinsic variations, and this would relax
the constraints on the distance of the partial coverer
from the emission region. 

If partial covering causes the observed variability, 
the viewing inclination would have to be such that our line
of sight intercepts the Compton thick 
occulting structure in the black 
hole region yet does not intercept a cold molecular torus. 
We do not have any firm evidence for a torus in \Irc, but
one seems plausible in light of its strong far-infrared emission.
The disc and torus axes could perhaps be misaligned.  
In addition, we note that one might expect to see
a reflection-dominated spectrum from cold matter when the direct 
flux is blocked. No such spectrum has been reported,
although it is possible to hide a reflection-dominated 
spectrum using scattering from highly ionized matter 
with a moderate Thomson depth. This scattering provides a 
separate light path which dilutes away the reflection-dominated
spectrum from the cold matter.   

Partial covering by Compton thick
material would not naturally explain the strong soft X-ray
excess of \Ir unless it were linked with additional interesting
physics. For example, one could imagine partial covering by
a serrated thick disc around a black hole. An unusually
high Eddington fraction could lead to both a thick disc
and a strong soft X-ray excess, and this picture would be
along the lines of that presented in Boroson \& Green (1992a).
The optical line properties of \Ir could also perhaps be 
explained following the discussion in Boroson \& Green (1992a).
Alternatively, one could invoke relativistic anisotropic 
continuum effects such as those discussed in section 5.1.7. 
However, this would almost inevitably imply the variability 
boosting effects also discussed in section 5.1.7 (thus perhaps 
obviating the partial covering). 

If the partial covering scenario is correct then there may be a
significant number of Seyferts whose black hole regions are hidden 
by their discs but not by their tori. We are not
aware of any objects that obviously fit this description. We also
note that the existence of some Compton thin Seyfert 2 galaxies 
probably rules out the ubiquitous presence of Compton thick
accretion disc elevations of the type described above.  


\subsubsection{Magnetic reconnection, shocks and pair instabilities in the
accretion disc corona}

The soft excess emission from \Ir 
that dominates its HRI count rate appears to
be roughly thermal in character when observed with the \ROSAT PSPC
and {\it ASCA\/}. Magnetic reconnection, 
shocks and pair instabilities in an
optically thin accretion disc corona would probably have trouble
producing such a spectrum and thus probably do not make the
bulk of the variability we are monitoring (see Abramowicz 1986 for
further discussion). If any of these did 
make the variability, we note that this would 
require a serious departure from the usual interpretation of the soft 
excess. 


\subsubsection{Accretion rate variations}

Accretion rate changes in a thin disc occur on 
the radial drift timescale,
as defined by equations 5.11 and 5.63 of 
Frank, King \& Raine (1992). Using reasonable estimates for
the black hole mass, accretion rate and viscosity parameter, 
we find that the radial drift timescale near the black
hole is much longer (by over a factor of 100) than 
the timescale over which giant amplitude variations are 
observed (see Figure 2). Thus accretion rate variations
do not seem to be a likely explanation for the observed
variability if the disc is thin. In addition, stable 
accretion in a disc tends 
to diffusively smooth out inhomogeneities (see figure 19 of
Frank, King \& Raine 1992) so it would be hard to create
the large accretion fluctuations required. 

It is perhaps possible that \Ir is accreting in an unstable
manner and that its accretion disc has large inhomogeneities 
due to the instability. While a detailed examination of this 
possibility is beyond the scope of this paper, we note that the
argument regarding the radial drift timescale is still likely
to be relevant at some level. Of course, the radial drift
timescale can be shortened if the Eddington fraction
is so high that the thin disc approximation breaks down.     


\subsubsection{Soft-excess temperature changes}

As described above, it appears that emission with a roughly thermal 
shape dominates the \ROSAT band. Changes 
in the temperature of a thermal component could lead to significant 
changes in count rate. Both real luminosity changes as well as
finite bandpass effects can occur
(i.e. flux can be added to and removed from the finite HRI band;
see the HRI response function illustrated in figure 11 of 
David et~al. 1996). From the fact that the radial drift 
timescale for a thin disc is much more 
than the observed giant variability timescale (see above), blackbody 
temperature changes due to accretion rate variations seem 
unlikely if the disc is thin. Blackbody temperature changes could also 
perhaps occur on the timescale over which sound waves propagate across
the emission region if they were due to large-scale thermal accretion disc
instabilities. If the inner disc is radiation pressure dominated then the 
sound crossing timescale is consistent with the observed 
variability timescale. 

As some types of extreme soft excess temperature 
changes appear possible in 
principle, we examine if they are consistent with the
available data. We have first estimated the amount by which 
the soft excess temperature would have to change to cause the
observed changes in HRI count rate. While the precise temperature
change required depends on the spectral model adopted, we find 
that a blackbody soft excess would have to drop in
temperature by a factor of $\simgt 2.5$. If the mean blackbody 
soft excess temperature of \Ir does persistently vary by a factor
of $\simgt 2.5$ due to large-scale thermal instabilities propagating
across the emission region, this would be quite remarkable. 
Such behaviour has not been seen from other Seyferts; the best
soft-excess monitoring data available appear to be consistent 
with a model in which the soft excess is constant in shape when 
it varies in flux (e.g. Done et~al. 1995). As noted
before, the soft X-ray luminosity of \Ir would be varying 
from that of a typical Seyfert to almost that of a quasar. 
We consider extreme soft excess
temperature changes to be fairly unlikely for \Ir due to
the fact that the \ROSAT PSPC and \ASCA have not recorded
such extreme spectral variations of this type
(note that pure finite bandpass effects would probably require
even stronger spectral variability). In fact, 
Otani (1995) and Otani et~al. (1996) state that the blackbody 
temperature they measure appears to be almost constant 
(to within about 15 per cent)
despite large \ASCA count rate variations. The 2--10 keV 
flux from \Ir has also been seen to 
be highly variable (Otani 1995), and in
this energy range the power law dominates the spectrum
rather than the soft excess. 


\subsubsection{Relativistic effects}

The rapid observed variations suggest that the X-ray emission 
we are monitoring originates close to the supermassive
black hole. In this region, the emitting particles will be
moving with relativistic bulk velocities, and an inevitable 
consequence of these velocities is that strong Doppler boosting  
will occur along many lines of sight
(e.g. Cunningham \& Bardeen 1973; Sunyaev 1973; section 4 of 
Guilbert, Fabian \& Rees 1983). A powerful relativistic jet 
preferentially pointed in the direction of Earth
seems unlikely from \Ir due to its radio properties
(see Section 3) and its position along the primary
Boroson \& Green (1992b) eigenvector, although we  
cannot formally rule out some type of X-ray jet emission.
Indeed, the nonlinear variability shown in Figure 1 is eerily
similar to the variability seen from blazars such as
3C~345. However, until such time as data 
necessitate the consideration of X-ray jet emission, we 
shall try to remain within the standard radio-quiet Seyfert 
picture (see Mushotzky, Done \& Pounds 1993).

Relativistic motions in an accretion disc will lead
to strong Doppler boosting. In addition, 
gravitational lensing by the black hole will
be important (e.g. Cunningham 1975; 
Rauch \& Blandford 1994). These
effects can lead to strong apparent flux variations
if the emission region is not steady and homogeneous, with 
the precise magnitude of such effects depending on the
location of the emission region, the viewing angle
and the degree of intrinsic variability or inhomogeneity. To 
illustrate this, we plot the `flux variability boost factor'
(hereafter the `boost factor') from 
section 4 of Guilbert, Fabian \& Rees (1983) versus the 
$\beta$ parameter (the velocity divided by $c$) for 
motion entirely along the line of sight (Figure 8).
The Doppler factor, $\delta=[\gamma(1-\beta \sin i)]^{-1}$, translates 
into a boost factor for variability ($dF/dt$ integrated over
a finite band) according to $\delta^{3+\Gamma}$ 
[here $\gamma=(1-\beta^2)^{-{1\over 2}}$ and $i$ is the disc 
inclination angle]. We show the results for three different values 
of photon index $\Gamma$ (while the precise spectral shape is 
not a power-law this works well for the qualitative illustration 
purposes relevant here). The boost factor can be remarkably large even 
for moderate $\beta$ values, and we note that these $\beta$
values will be achieved over a substantial region of the
inner accretion disc.   
    
\begin{figure}
       \psfig{figure=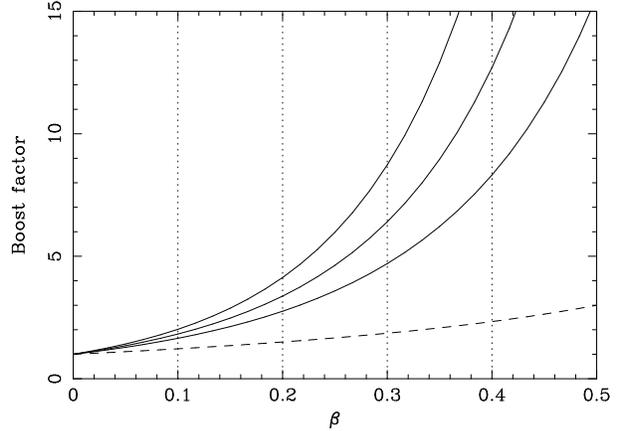,width=8.7cm,angle=270}
      \caption{
Flux variability boost factor 
from Guilbert et~al. (1983) versus $\beta$
for motion entirely along the line of sight. The curves 
are, from bottom to top, for photon indices ($\Gamma$) 
of 2, 3 and 4. Note the large boost factors even for 
moderate $\beta$ values. The dashed curve shows the ratio
of the boost factor for $\Gamma=4$ to that for $\Gamma=2$.  
}
\end{figure}

We shall first consider the possibility that the giant variability
of \Ir is due only to the `usual' Seyfert relativistic effects being
applied to an unusually steep spectrum ($\Gamma\approx 4$
rather than $\Gamma\approx 2$ if one approximates the spectrum
with a simple power-law model). This possibility can be roughly
examined using the dashed line in Figure 8, which shows the
ratio of the boost factor for $\Gamma=4$ to that for $\Gamma=2$. 
This ratio is not large for the values of $\beta$ shown in 
our plot, and it does not reach the required 
value of $\simgt 10$ until $\beta\approx 0.8$. 
Thus, unless most of the \ROSAT band emission from Seyferts originates
{\it extremely\/} close to their black holes (within $\sim 2$ Schwarzschild
radii), it appears that we are not just seeing the `usual' Seyfert
relativistic effects being applied to an unusually steep spectrum. 
Rather, it appears that we are probably observing unusually
strong relativistic effects. 

Unusually strong relativistic effects could be observed from 
\Ir if the emission from its disc were concentrated
closer to the black hole than usual or if its disc were
highly inclined. The first possibility could occur
if the black hole in \Ir were spinning faster than usual 
and thus its accretion disc extended closer to the hole.
One could speculate, parallel to the lines of 
Wilson \& Colbert (1995), that appropriate types of mergers lead
to fast spinning black holes in NLS1 objects, and  
Halpern \& Oke (1987) and Gaskell \& Koratkar (1997) have 
suggested that interactions 
may be especially important in triggering
NLS1 type activity. Systematic studies are needed before
such issues can be seriously discussed. We note that 
if the `spin paradigm' for radio-quiet/radio-loud
unification is true then the radio-quiet nature of \Ir (and its
low $R$ value; see Section~3) would perhaps be hard to understand 
if its black hole were spinning faster than usual for a 
Seyfert. This is not to say that we think the black 
holes in \Ir and other Seyferts are not rapidly spinning Kerr
black holes, but just that schemes which 
propose {\it more\/} rapidly spinning holes 
in NLS1 (cf. section 4 of Forster \& Halpern 1996) may 
come into conflict with the `spin paradigm.' It is 
worth noting that NLS1 objects like \Ir are the ones 
`furthest away' from radio-loud objects along the primary 
Boroson \& Green (1992b) eigenvector (see their figure 3). 
Schemes which propose that more rapidly spinning black
holes are the root cause of NLS1 characteristics (and
thus that black hole spin drives the Boroson \& Green 1992b 
eigenvector) would have to explain this fact. 

The second possibility mentioned above, a highly inclined 
accretion disc, leads to an `edge-on' model for \Irc. In
Figure 9 we plot boost factors at $5R_{\rm s}$ versus inclination for a
disc around a Schwarzschild black hole [following 
Paczy\'nski \& Wiita 1980 the velocity was calculated
assuming the Newtonian pseudo-potential $GM/(R-R_{\rm s}$);
we use a Schwarzschild black hole for illustrative purposes
because the calculations are more straightforward]. For a 
typical Seyfert 1 galaxy for which
$\Gamma\approx 2$ and the inclination is $\sim 30$ deg, the boost factor is
about two, whereas a source with a steep spectrum of
$\Gamma\approx 4$ (like \Irc) observed at an inclination of 
say 75 deg has a boost factor of about 16. 

If the variations of \Ir are due to the relativistic
amplification of emission regions on a disc, we can estimate 
a lower limit to the mass of the central black hole.
While this limit is rough and 
model-dependent, it also serves as
a useful and illustrative consistency check. Assuming 
from Figure~9 that the emission region under amplification
is within $5R_{\rm s}$ in order that there
is a large enough boost factor, changes by a factor of 2 occur for a
change of viewing angle of about 20 deg. Thus the fast doubling time
of $\sim 1000$~s from Section 2.3 (and PSPC observations) 
corresponds to an orbital period of $\sim 1.8\times 10^4$~s 
(here we have assumed that the emission region remains constant
during the time it takes to revolve through these 20 deg; 
if there are also intrinsic variations of the region then the 
period can be longer). From the derived radius and
period we then obtain a lower limit of $2\times 10^7$~M$_\odot$ for 
the black hole mass. This is an entirely reasonable number
suggesting that this possibility is at least plausible. 
We do not detect a strong peak in our Lomb-Scargle periodogram
near $5.6\times 10^{-5}$ Hz so if active regions do remain
roughly constant during the time it takes to revolve through
20 deg they probably do not remain constant throughout an
entire orbit. 

\begin{figure}
       \psfig{figure=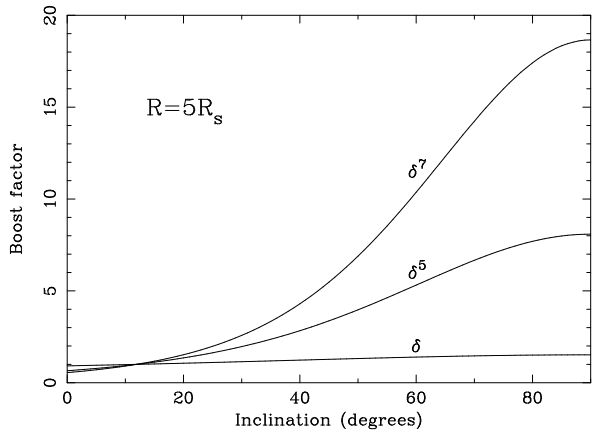,width=8.7cm,angle=270}
      \caption{
Boost factors versus inclination at $5R_{\rm s}$. 
The `$\delta$' curve shows the Doppler factor, 
the `$\delta^5$' curve shows the variability 
boost factor for $\Gamma=2$, and
the `$\delta^7$' curve shows the variability
boost factor for $\Gamma=4$.
}
\end{figure}

We note that a disc around a Kerr black hole observed at 
high inclination could account for both the remarkable 
variability, as discussed above, and for the 
strong soft X-ray excess. The beaming of radiation
into the equatorial plane of the disc, both due to the motion of the
emitting matter and gravitational light bending,
causes the luminous hot innermost regions around the black hole to be
best observed at high disc inclinations (see Cunningham 1975). 
Sun \& Malkan (1989) have calculated spectra of accretion discs in 
this case and it is clear that the soft X-ray excess in the spectrum 
is much greater for a disc viewed edge on than face on.
The very peculiar spectral features reported by Otani (1995)
may arise from a complex relativistically-boosted 
reflection spectrum from the disc. 

Despite the elegant success of the edge-on model in explaining the 
steep X-ray spectrum and large X-ray variability 
of \Ir (note it could also explain the similar behaviour seen
from PKS~0558--504 and PHL~1092), we must
admit that there currently appear to be difficulties 
in extending this model to the general optical
properties of NLS1. The low [O {\sc iii}] isotropic luminosities 
in NLS1, for example, argue that orientation effects alone cannot 
drive the Boroson \& Green (1992b) eigenvector (see Boroson 1992).
Furthermore, even if one managed to 
evade the strong arguments of Boroson (1992), 
the narrow permitted lines would presumably imply a polar outflow or 
inflow for the broad line clouds, contrary to the motion that has been 
suggested at least for radio-loud objects (where the disc orientation 
is assumed to be known from radio-jet observations; see the references
in Boroson 1992). An edge-on model does not obviously explain 
the strong Fe~{\sc ii} from \Ir either. Finally, 
our line of sight must not intercept any torus in this strong
IRAS emitter (the disc and torus might not be aligned). 
Further research is needed to examine these issues. 

One method by which the arguments of Boroson (1992) might be 
able to be escaped is if [O~{\sc iii}] is obscured in NLS1
by matter in their equatorial planes 
(see Hes, Barthel \& Fosbury 1993). However,
[O~{\sc iii}] would have to be quite highly anisotropic
and there is currently no clear evidence that the situation is
this extreme. In any case, we note that following this line of 
argument independently leads to an edge-on orientation. Tests of
Boroson (1992) using [O~{\sc ii}] would be helpful for
addressing this matter. 

\subsection{General implications and future investigations}

\subsubsection{Comparison with ultrasoft Galactic black hole candidates}

RE~J~1034+393, another ultrasoft narrow-line Seyfert~1 galaxy, has 
been noted to bear a striking similarity 
to the ultrasoft high states of Galactic
black hole candidates (Pounds, Done \& Osborne 1996). In 
particular, 
(1) its spectrum is dominated in the \ROSAT band by a giant soft X-ray excess, 
(2) its 2--10 keV photon index is an anomalously steep $\approx 2.6$, and
(3) it has not shown strong X-ray variability. 
The final characteristic is relevant because the ultrasoft components of 
Galactic black hole candidates show markedly reduced variability from 
the usual millisecond `flickering' that is observed, although they
do vary on timescales of about a day. One might expect the soft
excess components of supermassive analogues to ultrasoft Galactic
black holes to be stable on timescales of thousands of
years if the variability timescale depends roughly on the black 
hole mass. This expectation is not realized for \Irc, where stronger
than usual rather than weaker than usual variability is observed.
The reason for this discrepancy is not clear, although it could
perhaps be explained if \Ir suffered from partial covering
by a thick disc but RE~J~1034+393 did not
(see section 5.1.3; RE~J~1034+393 would also have
a high Eddington fraction but it would be aligned
so that our line of sight does not intercept the edge
of the thick disc). There might also be
significant heterogeneity among the ultrasoft NLS1. 

\subsubsection{Relativistic effects and steep spectrum Seyferts}

In light of the discussion above about relativistic 
effects, it is worth considering whether the other giant
amplitude variability events seen from ultrasoft NLS1 could  
have similar origins. For WPVS007 and PHL 1092, none of the
currently published data appears to rule out this possibility.
For Zwicky 159.034, however, the remarkable optical line changes
observed after the giant X-ray `outburst' appear to support a
real change in the energy generation rate of the 
central engine of this galaxy, or at least a drastic
repartitioning of its spectral energy distribution 
(see figure 4 of Pounds \& Brandt 1997 and 
the associated discussion).  
It is interesting and perhaps relevant that \Irc, WPVS007
and PHL 1092 are all strong Fe {\sc ii} emitters while
the main optical iron line emission from Zwicky 159.034 
(and RE~J~1034+393) was from high ionization forbidden 
iron lines. 
 
As mentioned above, Green, McHardy \& Lehto (1993) have found
evidence that sources with the steepest (1--8~keV) energy spectra
are possibly the most variable. They propose that this may be 
due to the fact that sources with flatter energy spectra have 
stronger Compton reflection humps. The reflection process is
thought to smooth out rapid variability. While this 
possibility seems plausible, we point out 
that sources with steeper energy spectra will also show 
stronger Doppler boosting when observed in a fixed energy
band (see Figure 8 and the associated text). This effect,
which is inevitable along many lines of sight, may also 
contribute to the correlation found by 
Green, McHardy \& Lehto (1993). The accretion 
process in Seyferts is
probably much more intrinsically stable than 
relativistically-boosted Seyfert light 
curves suggest at first inspection. 
We also comment that if the discs and tori of Seyferts
are aligned (as suggested by observations of ionization
cones), the high energy direct emission from 
Compton thin Seyfert 2s should suffer from particularly 
strong Doppler boosting effects (see Figure 9).  

\subsubsection{Future investigations}

In light of these first NLS1 monitoring results additional 
systematic X-ray monitoring of ultrasoft NLS1 is needed. 
We are planning to build upon this work by making 
\ROSAT HRI monitoring observations of PHL~1092 and
other carefully selected ultrasoft NLS1. 
One of the goals is to determine whether ultrasoft 
NLS1 generally show stronger nonlinearity than is
seen in more typical Seyfert~1s. Stronger nonlinearity
could suggest that we are seeing an additional,
previously unrecognized, nonlinear X-ray emission process 
in NLS1. Future monitoring 
by instruments with good spectral resolution and
sensitivity in both the soft and hard X-ray 
bands would be ideal, as spectral variability could then be
rigorously constrained and studied. For example, it would
be interesting to perform a precision search for spectral
changes during nonlinear variability events, as these
could help to determine whether there is more than one
X-ray emission process. 

Monitoring of \Ir at other wavelengths would also be helpful.
If the observed X-ray variability is due to real changes in
the energy generation rate of the nucleus, then one might
expect strong changes at other wavelengths as well. On
the other hand, if the X-ray variability is due to 
relativistic amplification or occulting structures in the
inner accretion disk, then one might not expect variability
that is stronger than usual. The remarkable Ly$\alpha$ line
core variations of \Ir need further study using HST, and
it would be also useful to reverberation map this strong
Fe~{\sc ii} emitter.

\section{Summary}

We have used the \ROSAT HRI to monitor the radio-quiet,
ultrasoft, strong Fe~{\sc ii}, narrow-line Seyfert~1
galaxy \Ir for 30 days. This is the first systematic
monitoring of an ultrasoft narrow-line Seyfert~1.
Our main results are the following:

(1) We have discovered persistent giant amplitude and rapid
HRI count rate variability from \Irc. Over the course of our
observations, we detect at least five giant amplitude 
variations. The maximum observed amplitude of variability
is about a factor of 60, and we detect a variation by about
a factor of 57 in just two days. 

(2) The variability of \Ir is probably nonlinear in character, 
and we describe the consequences of this nonlinearity.

(3) Simple tests do not reveal any highly significant periodicity.
More sophisticated analyses of the light curve are in progress.

(4) We carefully examine the available data, and find no reason
to suspect the identification of the X-ray source.

(5) We examine possible explanations for the persistent giant
amplitude variability. 
Unusually strong relativistic effects
and partial covering by occulting structures on an accretion
disc provide plausible scenarios needing further 
investigation. An edge-on orientation of
the inner accretion disc could lead to both strong 
relativistic variability enhancement as well as a strong
soft X-ray excess. 
A small X-ray jet with nonlinear variability is another 
interesting possibility, although there is currently no 
hard evidence for such emission. 

(6) We discuss the relevance of these observations to the
previously suggested analogy between ultrasoft Galactic 
black hole candidates and ultrasoft Seyferts. We examine
whether relativistic boosting effects may be generally
relevant to ultrasoft Seyferts.

\section*{Acknowledgments}
We thank 
B. Aschenbach, 
A. Celotti, 
C. Done, 
M. Elvis,
A. Green, 
J. Halpern,
C. Otani,
A. Siemiginowska, 
R. Sunyaev,
J. Tr\"umper and
B. Wilkes
for helpful discussions.  
We thank C. Reynolds and M. Ward for showing us their optical 
spectra of \Ir prior to publication. 
We thank M. Irwin and R. McMahon for help with the optical
plates.  
The \ROSAT project is supported by the Bundesministerium
f\"ur Forschung und Technologie (BMFT) and the
Max-Planck-Society.
We gratefully acknowledge the support of the
Smithsonian Institution and the  
United States National Science Foundation (WNB)
and the United Kingdom Royal Society (ACF).  
  
{}

\bsp


\begin{thebibliography}{}

\bibitem []{} Abramowicz M.A., 1986, in Treves A., ed, Variability of
Galactic and Extragalactic X-ray Sources. Associazione per 
L'Avanzamento Dell'Astronomia, Milan, p. 137
%
\bibitem []{} Abramowicz M.A., Lanza A., Spiegel E.A., Szuszkiewicz E.,
1992, Nature, 356, 41
%
\bibitem []{} Bao G., 1992, A\&A, 257, 594
%
\bibitem []{} Bevington P.R., 1969, Data Reduction and Error Analysis
for the Physical Sciences. McGraw-Hill, New York
%
\bibitem []{} Boller Th., Tr\"umper J., Molendi S., Fink H., Schaeidt S.,
Caulet A., Dennefeld M., 1993, A\&A, 279, 53
%
\bibitem []{} Boller Th., Brandt W.N., Fink H., 1996, A\&A, 305, 53 (BBF96)
%
\bibitem []{} Boroson T.A., Green R.F., 1992a, in 
Holt S.S., Neff S.G., Urry C.M., eds, Testing the AGN Paradigm.
AIP Press, New York, p. 584
%
\bibitem []{} Boroson T.A., Green R.F., 1992b, ApJS, 80, 109
%
\bibitem []{} Brandt W.N., 1995, PhD thesis, University of Cambridge 
%
\bibitem []{} Brandt W.N., Pounds K.A., Fink H., 1995, MNRAS, 273, 47P
%
\bibitem []{} Brandt W.N., Pounds K.A., Fink H., Fabian A.C., 1996a, 
in Zimmermann H.U., Tr\"umper J.E., Yorke H., eds, R\"ontgenstrahlung 
from the Universe. MPE Press, Garching, p. 429 (MPE Report 263)
%
\bibitem []{} Brandt W.N., Fabian A.C., Dotani T., Nagase F., Inoue H., 
Kotani T., Segawa Y., 1996b, MNRAS, 283, 1071
%
%
\bibitem []{} Cruddace R.G., Hasinger G.R, Schmitt J.H.M.M., 1988, in
Murtagh F., Heck A., eds, Astronomy from Large Databases: Scientific
Objectives and Methodological Approaches. ESO Press, Garching, p. 177
(ESO Conference and Workshop Proceedings 28)
%
\bibitem []{} Cunningham C.T., Bardeen J.M., 1973, ApJ, 183, 237
%
\bibitem []{} Cunningham C.T., 1975, ApJ, 202, 788
%
\bibitem []{} David L.P., Harnden F.R., Kearns K.E., Zombeck M.V., 1996, 
The \ROSAT High Resolution Imager. SAO Press, Cambridge
%
\bibitem []{} Done C., Madejski G.M., Mushotzky R.F., Turner T.J., 
Koyama K., Kunieda H., 1992, ApJ, 400, 138
%
\bibitem []{} Done C., Pounds K.A., Nandra K., Fabian A.C., 1995, 
MNRAS, 275, 417
%
\bibitem []{} Eadie W.T., Drijard D., James F.E., Roos M., Sadoulet B.,
1971, Statistical Methods in Experimental Physics. North-Holland, Amsterdam
%
\bibitem []{} Elvis M., 1976, MNRAS, 177, L7
%
%
\bibitem []{} Fabian A.C., 1979, Proc. Roy. Soc. London A, 366, 449
%
\bibitem []{} Falcke H., Patnaik A.R., Sherwood W., 1996, ApJ, 473, L13
%
\bibitem []{} Forster K., Halpern J.P., 1996, ApJ, 468, 565 
%
\bibitem []{} Frank J., King A., Raine D., 1992, Accretion Power in 
Astrophysics. Cambridge Univeristy Press, Cambridge
%
\bibitem []{} Gaskell C.M., Koratkar A., 1997, ApJ, submitted
%
\bibitem []{} Gondhalekar P.M., Kellett B.J., Pounds K.A., Matthews L.,
Quenby J.J., 1994, MNRAS, 268, 973
%
\bibitem []{} Green A.R., McHardy I.M., Lehto H.J., 1993, MNRAS, 265, 664
%
%
\bibitem []{} Green A.R., 1993, PhD thesis, University of Southampton
(http://sousun1.phys.soton.ac.uk/pubs/Publications.html)
%
\bibitem []{} Grupe D., Beuermann K., Mannheim K., Bade N., Thomas
H.-C., de Martino D., Schwope A., 1995a, A\&A, 299, L5
%
\bibitem []{} Grupe D., Beuermann K., Mannheim K., Thomas H.-C., Fink
H.H., 1995b, A\&A, 300, L21 
%
\bibitem []{} Guilbert P.W., Fabian A.C., Rees M.J., 1983, MNRAS, 205, 593
%
\bibitem []{} Halpern J.P., Oke J.B., 1987, ApJ, 312, 91
%
\bibitem []{} Hes R., Barthel P.D., Fosbury R.A.E., 1993, 
Nature, 362, 326
%
\bibitem []{} Hook I.M., McMahon R.G., Boyle B.J., Irwin M.J., 1994,
MNRAS, 268, 305
%
\bibitem []{} Jenkner H., Lasker B., Sturch C., McLean B., Shara M., 
Russell J., 1990, AJ, 99, 2081
%
\bibitem []{} Kellermann K.I., Sramek R., Schmidt M., Shaffer D.B.,
Green R., 1989, AJ, 98, 1195
%
\bibitem []{} Kinman T.D., 1968, Science, 162, 1081
%
\bibitem []{} Lasker B., Sturch C., McLean B., Russell J., 
Jenkner H., Shara M., 1990, AJ, 99, 2019 
%
%
\bibitem []{} Leighly K.M., Marshall H., 1996, 
The \ROSAT Newsletter: Special Edition, 13, 46
%
\bibitem []{} Lightman A., Giacconi R., Tananbaum H., 1978, ApJ, 224, 375
%
\bibitem []{} Maccacaro T., Gioia I.M., Wolter A., Zamorani G., 
Stocke J.T., 1988, ApJ, 326, 680
%
\bibitem []{} Mas-Hesse J.M., Rodr\' \i guez-Pascual P.M., 
Sanz Fern\' andez de C\' ordoba L., Boller Th., 1994, A\&A, 283, L9
%
\bibitem []{} McHardy I.M., Green A.R., Done C., Puchnarewicz E.M.,
Mason K.O., Branduardi-Raymont G., Jones M.H., 1995, MNRAS, 273, 549
%
\bibitem []{} Morse J.A., 1994, PASP, 106, 675 
%
\bibitem []{} Mushotzky R.F., Done C., Pounds K.A., 1993, ARAA, 31, 717
%
%
\bibitem []{} Narayan R., Bartelmann M., 1997, in
Dekel A., Ostriker J.P., eds, Formation of Structure in the
Universe. Cambridge Univ. Press, Cambridge, in press
%
\bibitem []{} Norris R.P., Allen D.A., Sramek R.A., Kesteven M.J.,
Troup E.R., 1990, ApJ, 359, 291
%
\bibitem []{} Otani C., 1995, PhD thesis, University of Tokyo
%
\bibitem []{} Otani C., Kii T., Miya K., 1996, in Zimmermann H.U.,
Tr\"umper J.E., Yorke H., eds, R\"ontgenstrahlung from the Universe.
MPE Press, Garching, p. 491 (MPE Report 263)
%
\bibitem []{} Paczy\'nski B., Wiita P.J., 1980, A\&A, 88, 23
%
\bibitem []{} Press W.H, Teukolsky S.A., Vetterling W.T., Flannery B.P.,
1992, Numerical Recipes in FORTRAN: Second Edition. Cambridge Univ. 
Press, Cambridge  
%
\bibitem []{} Pounds K.A., Done C., Osborne J., 1995, MNRAS, 277, L5
%
\bibitem []{} Pounds K.A., Brandt W.N., 1997, in 
Makino F., Mitsuda K., eds,
X-Ray Imaging and Spectroscopy of Cosmic Hot Plasmas: \ASCA Third 
Anniversary Proceedings. Univ. Acad. Press, Tokyo, in press
%
\bibitem []{} Puchnarewicz E.M., Mason K.O., Siemiginowska A., Pounds K.A.,
1995, MNRAS, 276, 20
%
\bibitem []{} Rauch K.P., Blandford R.D., 1994, ApJ, 421, 46
%
\bibitem []{} Remillard R.A., Grossan B., Bradt H.V., Ohahsi T.,
Hayashida K., Makino F., Tanaka Y., 1991, Nature, 350, 589
%
\bibitem []{} Russell J., Lasker B., McLean B., Sturch C., Jenkner H.,
1990, AJ, 99, 2059.
%
\bibitem []{} Sandage A., Tammann G.A., 1987, A Revised Shapley-Ames Catalog
of Bright Galaxies. Carnegie Institute of Washington Press, Washington D.C.
%
\bibitem []{} Scargle J.D., 1982, ApJ, 263, 835
%
\bibitem []{} Schmidt M., Green R.F., 1983, ApJ, 269, 352  
%
\bibitem []{} Schmidt M., Green R.F., 1986, ApJ, 305, 68
%
\bibitem []{} Schneider P., Weiss A., 1987, A\&A, 171, 49
%
\bibitem []{} Stark A.A., Gammie C.F., Wilson R.W., Bally J., Linke R., 
Heiles C., Hurwitz M., 1992, ApJS, 79, 77 
%
\bibitem []{} Sun W.-H., Malkan M.A., 1989, ApJ, 346, 68 
%
\bibitem []{} Sunyaev R.A., 1973, Soviet Astronomy AJ, 16, 941
%
\bibitem []{} Tananbaum H., Peters G., Forman W., Giacconi R., 
Jones C., Avni Y., 1978, ApJ, 223, 74
%
\bibitem []{} Terrell J., 1967, ApJ, 147, 827
%
\bibitem []{} Thorne K.S., 1974, ApJ, 191, 507
%
\bibitem []{} Tr\"umper J., 1983, Adv. Space Res., 4, 241
%
\bibitem []{} Vio R., Cristiani S., Lessi O., Salvadori L., 1991, ApJ,
380, 351 
%
\bibitem []{} Vio R., Cristiani S., Lessi O., Provenzale A., 1992, ApJ, 
391, 518
%
\bibitem []{} Wilson A.S., Colbert E.J.M., 1995, ApJ, 438, 62
%
\bibitem []{} Zwicky F., 1971, Catalogue of Selected Compact Galaxies and of
Post-Eruptive Galaxies. Offsetdruck L. Speich Zuerich, Zuerich


\end{thebibliography}
\end{document}